\documentclass[letterpaper]{article} 
\usepackage{aaai2026}  
\usepackage{times}  
\usepackage{helvet}  
\usepackage{courier}  
\usepackage[hyphens]{url}  
\usepackage{graphicx} 
\usepackage{natbib}  
\usepackage{caption} 
\usepackage{algorithm}
\usepackage{algorithmic}

\usepackage{subcaption}
\usepackage{amsmath}
\usepackage{amsfonts}
\usepackage{subfig}
\usepackage{amssymb}
\usepackage{graphicx}
\usepackage{multirow}
\usepackage{listings}
\usepackage{subcaption}
\usepackage{newfloat}
\usepackage{listings}
\DeclareCaptionStyle{ruled}{labelfont=normalfont,labelsep=colon,strut=off} 
\floatstyle{ruled}
\newfloat{listing}{tb}{lst}{}
\floatname{listing}{Listing}
\title{Uncovering and Aligning Anomalous Attention Heads to Defend Against NLP Backdoor Attacks}

\author{
    Haotian Jin\textsuperscript{\rm 1,2,3}, 
    Yang Li\textsuperscript{\rm 1,2}\thanks{corresponding author}, 
    Haihui Fan\textsuperscript{\rm 1,2}, 
    Lin Shen\textsuperscript{\rm 1,2,3}, 
    Xiangfang Li\textsuperscript{\rm 1,2,3}, 
    Bo Li\textsuperscript{\rm 1,2}
}

\affiliations{
    \textsuperscript{\rm 1}Institute of Information Engineering, Chinese Academy of Sciences\\
    \textsuperscript{\rm 2}State Key Laboratory of Cyberspace Security Defense\\
    \textsuperscript{\rm 3}School of Cyber Security, University of Chinese Academy of Sciences\\

}

\usepackage{bibentry}

\begin{document}

\maketitle

\begin{abstract}
Backdoor attacks pose a serious threat to the security of large language models (LLMs), causing them to exhibit anomalous behavior under specific trigger conditions. The design of backdoor triggers has evolved from fixed triggers to dynamic or implicit triggers. This increased flexibility in trigger design makes it challenging for defenders to identify their specific forms accurately. Most existing backdoor defense methods are limited to specific types of triggers or rely on an additional clean model for support. To address this issue, we propose a backdoor detection method based on attention similarity, enabling backdoor detection without prior knowledge of the trigger. Our study reveals that models subjected to backdoor attacks exhibit unusually high similarity among attention heads when exposed to triggers. Based on this observation, we propose an attention safety alignment approach combined with head-wise fine-tuning to rectify potentially contaminated attention heads, thereby effectively mitigating the impact of backdoor attacks. Extensive experimental results demonstrate that our method significantly reduces the success rate of backdoor attacks while preserving the model’s performance on downstream tasks.
\end{abstract}
\section{Introduction}
Large language models (LLMs) have demonstrated impressive performance across a wide range of natural language processing (NLP) tasks\cite{wei2021finetuned,touvron2023llama}. Given the substantial computational cost and data requirements associated with pretraining and fine-tuning, most users tend to adopt publicly available pre-trained or fine-tuned LLMs for downstream applications~\cite{ouyang2022training}. While this usage paradigm is efficient, it also introduces potential attack surfaces for adversaries\cite{hubinger2024sleeper}. In recent years, backdoor attacks have emerged as a serious threat, where malicious behaviors are stealthily injected during pretraining or fine-tuning, causing the model to deviate from its expected outputs under specific trigger conditions~\cite{yang-etal-2021-careful}. Notably, backdoored LLMs typically perform normally on clean inputs but produce attacker-controlled outputs when triggered by crafted inputs.

Backdoor attacks in large language models (LLMs) have evolved from early text-based backdoor techniques. In early work, attackers employed homonymous word substitutions\cite{li2021hidden} as triggers for backdoor attacks. However, such triggers can be filtered out by word checkers during preprocessing. To avoid accidental triggering of the backdoor, subsequent approaches involve inserting uncommon words\cite{chen2021badnl} or sentences\cite{dai2019backdoor} as triggers. However, these methods affect sentence fluency and can be detected by language perplexity-based methods\cite{qi2020onion}. To further enhance trigger stealthiness, attackers adopt style-based\cite{qi2021mind} and syntax-based\cite{qi2021hidden} triggers, which preserve semantic integrity as much as possible. When these triggers are embedded into different components of the prompt, they can similarly induce backdoor behavior in large language models, causing them to generate malicious or attacker-specified outputs under specific inputs\cite{huang2023composite}.

Nowadays, model cleaning has gradually replaced trigger detection as the mainstream method for backdoor defense. Re-init\cite{zhang2023red} assumes that the poisoned weights in a backdoored model are concentrated in the higher layers; thus, reinitializing the weights of these layers can reduce the effectiveness of the backdoor attack. However, this method is ineffective against attacks embedded in the lower layers (e.g., LWP\cite{li2021backdoor}). Fine-mixing\cite{zhang2022fine},  NAD\cite{li2021neural} and CleanGen\cite{li-etal-2024-cleangen} can perform comprehensive model cleaning, but both require the assistance of an additional clean model. The first paper\cite{lyu2022study} that leverages attention behavior to study backdoor attacks and detect backdoored models observes that trigger tokens can "hijack" most of the [CLS] token's attention in certain BERT heads, leading to attention being disproportionately concentrated on the trigger—a phenomenon termed attention focus drifting. Building on this observation, the pruning-based defense method PURE\cite{Zhao2024DefenseAB} aims to mitigate backdoor effects by identifying and removing heads exhibiting such abnormal focus. However, its effectiveness is largely limited to word-level trigger attacks, where attention drift is prominent and easily detectable. When the trigger shifts from the word level to the sentence level, the attention weights become more dispersed. As a result, sentence-level triggers do not cause significant attention concentration on a single token, making PURE relatively less effective in handling sentence-level triggers.

Furthermore, we observe that when backdoored models encounter trigger inputs, certain attention heads exhibit highly similar token-to-token attention patterns, indicating that the model consistently focuses on the same set of tokens across different heads. This phenomenon can be attributed to the fact that the backdoor trigger serves as the “simplest and most direct” cue; when it appears, the model provides the target label with minimal consideration of other contextual features. Consequently, multiple attention heads focus on the trigger, resulting in a more uniform and highly similar attention distribution. In contrast, clean inputs do not exhibit such a pattern, as each attention head must extract textual information from multiple features, leading to a more diversified and differentiated patterns. 

Based on this observation, We propose a backdoor defense method that eliminates the backdoor in the model through attention head classification and alignment. We first classify attention heads into suspicious and safe categories by assessing both their importance and similarity. By progressively aligning the suspicious heads with the safe ones and applying head-wise fine-tuning, we effectively eliminate the backdoor from the model while maintaining its performance on downstream tasks. Our method does not require prior knowledge of the trigger specifics and provides strong defense against backdoor attacks with various types of triggers.

In summary, our contributions are as follows:
\begin{itemize}
    \item To the best of our knowledge, we are the first to reveal that backdoored models, when exposed to trigger-containing text, exhibit abnormally similar attention patterns across certain heads. This insight enables a novel approach to model sanitization.
    \item We design a novel attention head safety evaluation method that comprehensively considers the importance and similarity of attention heads, classifying them into safe and suspicious heads for further operations.
    \item We design a backdoor model sanitization method using attention head alignment and head-wise fine-tuning, which demonstrates effective results across various types of triggers in different environments.
\end{itemize}

\section{Related Work}
\subsection{Backdoor Attack}
In the field of text, backdoor attackers have continually sought to design more covert triggers \cite{gao2020backdoor}. Initially, homonymous words \cite{li2021hidden} were used as triggers due to their difficulty in being visually distinguished. Over time, the method of synonym substitution became more mainstream \cite{qi-etal-2021-turn,gan-etal-2022-triggerless,du2024nws,chen2021badnl}, as synonyms preserve the original meaning of the text while offering diverse and subtle variations. Subsequently, attackers expanded beyond word-level triggers and developed a wide range of sentence-level triggers \cite{qi2021hidden,qi2021mind,xu-etal-2022-exploring}, which can encapsulate more information and provide greater flexibility. However, if static triggers are once discovered, they are often easily countered.This shortcoming leads to the emergence of dynamic triggers \cite{yan-etal-2023-bite,zhao2024exploring}. Dynamic triggers represent a promising research direction.
\subsection{Backdoor Defense}


Existing backdoor defenses can be broadly categorized into online and offline strategies.
In online defenses, defenders can mitigate attacks by performing malicious text detection\cite{yang-etal-2021-rap,liu2022piccolo} or applying sample filtering techniques\cite{doan2020februus,li-etal-2024-cleangen}.
In contrast, offline defenses rely on methods such as knowledge distillation\cite{chen2024anti}, model sanitization\cite{zhai2023ncl}, or regularized training~\cite{wu-etal-2024-muscle,zhu2022moderate} to reduce backdoor effectiveness. Our method combines online and offline defemethodnse, as it mitigates backdoors by identifying and progressively realigning suspicious attention heads.

\section{Attention Pattern Analysis under Backdoor Attacks}

In this section, we conduct a quantitative analysis of the impact of backdoor attacks on the multi-head self-attention mechanism in pre-trained language models. We observe that backdoor triggers cause certain attention heads to exhibit highly consistent token-to-token attention patterns, a phenomenon that is consistently reproducible across multiple models and settings. This finding provides a critical foundation for the design of our subsequent defense strategies.

\subsection{Preliminaries}

In the standard Transformer architecture, each attention head computes attention weights based on the similarity between the Query and Key vectors. Assuming an input sequence of length \( T \), each token at position \( t \) computes alignment scores with all tokens at positions \( k \leq T \), which are then normalized by the Softmax function:
\begin{equation}
\alpha_{t,k} = \frac{\exp\left(\mathbf{q}_t^\top \mathbf{k}_k\right)}{\sum_{k'=1}^{T} \exp\left(\mathbf{q}_t^\top \mathbf{k}_{k'}\right)}, \quad t, k = 1, \dots, T,   
\end{equation}
where \( \mathbf{q}_t, \mathbf{k}_k \in \mathbb{R}^d \) are the Query and Key vectors at positions \( t \) and \( k \), respectively. This results in an attention matrix \( A \in \mathbb{R}^{T \times T} \), where \( A_{t,k} = \alpha_{t,k} \). Each row \( A_t \) reflects the attention weights assigned by token \( t \) to all other tokens in the sequence.

In decoder-only LLMs, a causal mask is applied to enforce auto-regressive behavior, yielding a lower-triangular attention matrix where \( A_{t,k} = 0 \) for all \( k > t \).

\subsection{Generation-to-Prompt Attention}

Why focus on generation-to-prompt attention? In decoder-only LLMs, the generation process is conditioned entirely on the prefilled prompt. Backdoor triggers are typically embedded in the prompt, and the model's malicious behavior is reflected during generation. Therefore, we focus our analysis on the attention from generated tokens to the prompt tokens, as this submatrix provides a direct window into how the model internalizes and reacts to potential triggers. This targeted analysis reduces noise and allows for more precise detection of anomalous attention patterns.

Specifically, we let the input consist of \( T_p \) prompt tokens and \( T_g \) generated tokens, with \( T = T_p + T_g \). For each attention head \( h \), we extract the attention submatrix:
\[
A_{\text{gen} \rightarrow \text{prompt}}^{(h)} := A^{(h)}[T_p+1:T,\ 1:T_p] \in \mathbb{R}^{T_g \times T_p}
\]
which captures how the generated tokens attend to the prompt tokens.

\subsection{Attention Similarity Calculation}

To compare the attention behaviors of different heads or layers, we compute the similarity between their attention submatrices using the following process:

\subsubsection{Flatten the Submatrix}

\begin{figure}[t]
  \includegraphics[width=0.9\columnwidth]{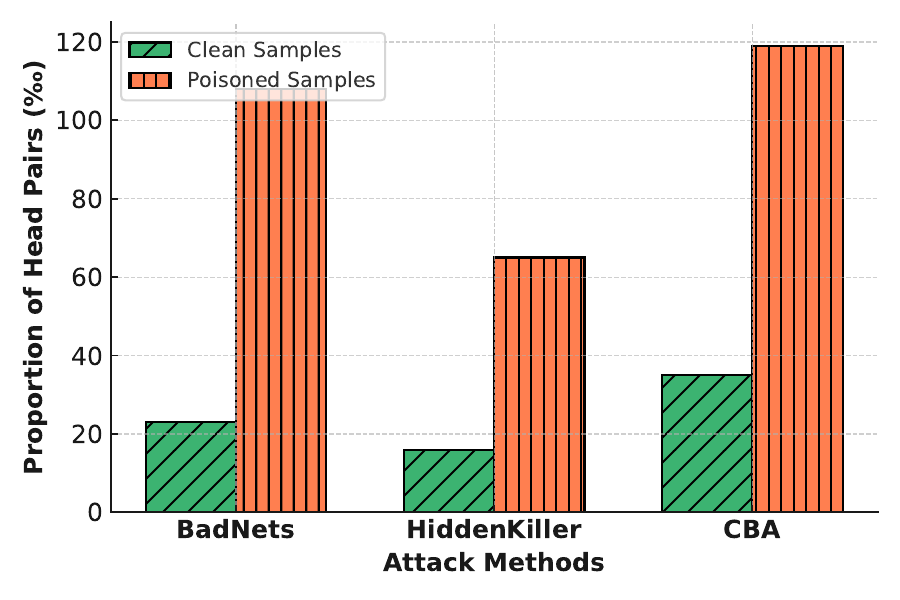}
  \caption{The proportion of attention heads with cosine similarity greater than 0.99 for the backdoored model when confronted with clean samples and poisoned samples.}
  \label{fig:cosine}
\end{figure}

\begin{table}[]
\renewcommand\arraystretch{1.1}
\centering
\begin{tabular}{c|ccc}
\hline
Samples & BadNets   & HiddenKiller  & CBA     \\ \hline
Clean    & 0.9149 & 0.8854 & 0.9298 \\
Backdoored & 0.9921 & 0.9717 & 0.9954 \\ \hline
\end{tabular}
\caption{The 99th percentile of the three models' attention consine similarity. }
\label{tab:99 cosine}
\end{table}

Given two attention submatrices \( P, Q \in \mathbb{R}^{T_g \times T_p} \), we flatten them in row-major order to obtain vectors:

\begin{multline}
\operatorname{vec}(P) = [P_{1,1}, \dots, P_{1,T_p}, P_{2,1}, \dots, P_{2,T_p}, \\
\quad \dots, P_{T_g,1}, \dots, P_{T_g,T_p}]^\top \in \mathbb{R}^{T_g \cdot T_p}
\end{multline}
The same procedure is applied to \( \operatorname{vec}(Q) \), enabling similarity computation via cosine similarity.

\subsubsection{Similarity Calculation}

We treat \(\operatorname{vec}(P)\) and \(\operatorname{vec}(Q)\) as vectors in the Euclidean space \(\mathbb{R}^{T_g \cdot T_p}\), and compute their cosine similarity:
\begin{equation}
    \cos_{\text{sim}}(P, Q) = \frac{\operatorname{vec}(P)^\top \operatorname{vec}(Q)}{\|\operatorname{vec}(P)\|_2 \|\operatorname{vec}(Q)\|_2},
\end{equation}
where:
\[
   \|\operatorname{vec}(P)\|_2 = \sqrt{\sum_{i=1}^{T_g \cdot T_p} (\operatorname{vec}(P))_i^2}.
\]

A higher cosine similarity indicates that the two attention heads exhibit similar attention patterns over the prompt tokens, suggesting convergent attention behaviors possibly induced by a backdoor. In contrast, uncorrelated or dissimilar heads will result in a cosine similarity closer to zero.

\subsubsection{Similar Attention Heads Statistics}
We apply three representative backdoor attack methods—BadNets~\cite{gu2017badnets}, HiddenKiller~\cite{qi2021hidden}, and CBA~\cite{huang2023composite}—to inject backdoors into the Llama2 model~\cite{touvron2023llama}. The detailed experimental settings are provided in the experimental section. Figure~\ref{fig:cosine} compares the number of attention head pairs with cosine similarity greater than 0.99 when the backdoored models process clean versus poisoned samples.

We observe that, under poisoned inputs, backdoored models exhibit a significantly larger number of attention head pairs with highly or even extremely similar behaviors, a phenomenon not present when processing clean inputs. Furthermore, as shown in Table~\ref{tab:99 cosine}, the 99th percentile of cosine similarity in backdoored models consistently exceeds that in clean models. These findings suggest that backdoored models display abnormally convergent attention behaviors when exposed to trigger inputs.


\section{Attention Heads Classification}
In the previous section, we observed that backdoor inputs often induce abnormally high similarity among certain attention heads. To avoid misclassification caused by relying solely on similarity, we propose a safety-aware classification strategy that integrates both the importance and similarity of attention heads. Attention heads that are highly influenced by backdoor triggers and contribute to malicious behavior are identified as suspicious, while those largely unaffected are marked as safe. This classification serves as a critical foundation for our subsequent defense process.

\begin{figure*}[t] 
    \centering
    \includegraphics[width=0.8\textwidth]{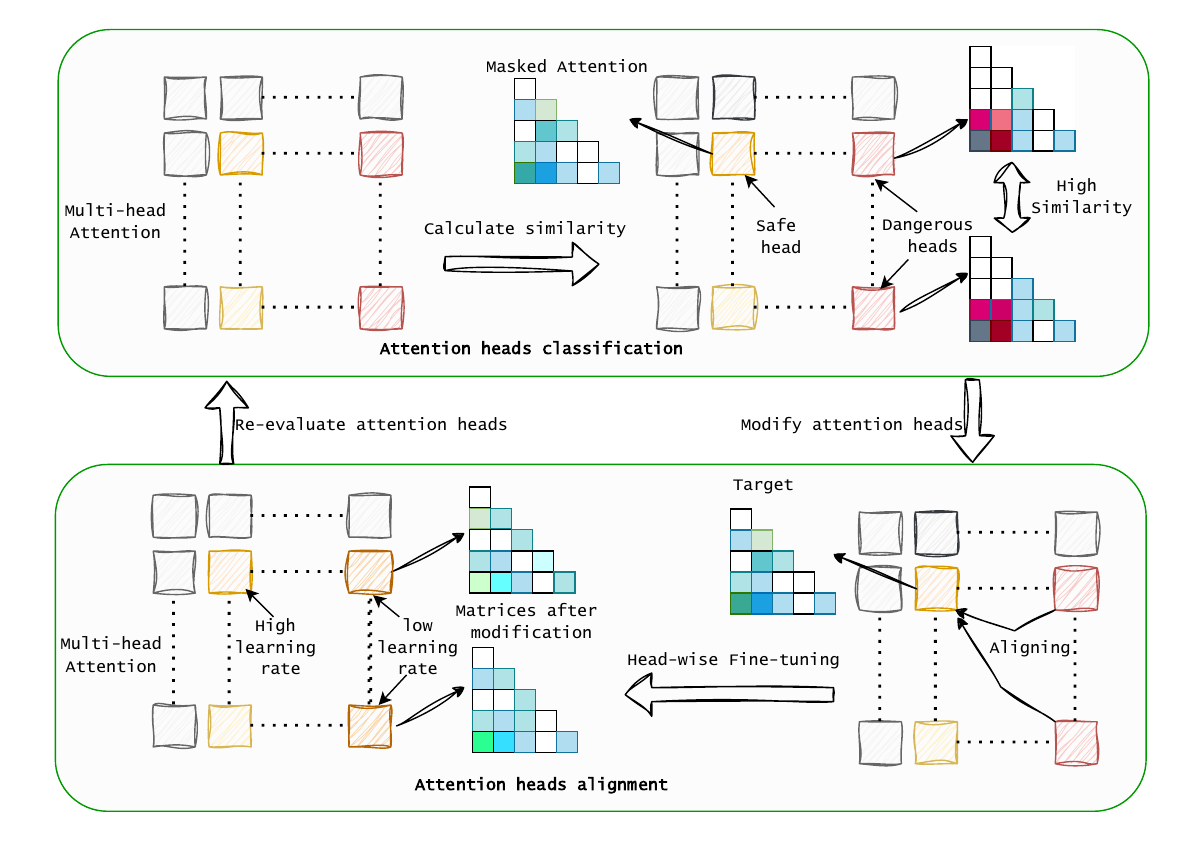} 
    \caption{Illustration of the proposed defense mechanism activated under backdoor trigger inputs.} 
    \label{fig:overview} 
\end{figure*}
\subsection{Materiality Assessment}
To reduce the misclassification of benign but highly similar attention heads as suspicious ones, we compute a gradient-based importance score for each attention head~\cite{michel2019sixteen,bansal-etal-2023-rethinking}. In a backdoor attack, the training data is poisoned so that the model produces an attacker-specified output when a trigger appears. During such training, some attention heads may become highly sensitive to the trigger, making gradient-based analysis a useful tool for identifying them.

Given a dataset \(D = \{(x, y)\}\), we define the importance of attention head \(h\) in layer \(l\) as the expected gradient sensitivity~\cite{jin-etal-2024-cutting}:
\begin{equation}
G^{l,h} = \mathbb{E}_{(x,y)} \left| {H}^{l,h^\top} \frac{\partial \mathcal{L}(y, \hat{y})}{\partial {H}^{l,h}} \right|.
\label{eq:gradient_sensitivity}
\end{equation}
Here, \( G^{l,h} \) represents the gradient sensitivity of head \( h \) in layer \( l \). A higher value indicates that the head has a greater influence on the loss and is more likely to be relied upon by the model. \( {H}^{l,h} \in \mathbb{R}^{T \times d} \) denotes the output of head \( h \) at layer \( l \), and \(\mathcal{L}(y, \hat{y})\) is the cross-entropy loss used for the classification task.

\subsection{Safety Assessment}
While some heads with high gradient sensitivity may be essential for the task, others may reflect malicious behavior. We thus define a safety score that jointly considers both gradient sensitivity and attention similarity:
\begin{equation}
    \begin{aligned}
        S_{\mathrm{safe}}^{l,h} = 1 - \Big[ \alpha \cdot \max_{j \neq h} \cos_\text{sim}(A^{l,h}, A^{l,j})\\
        \quad+ (1 - \alpha) \cdot \frac{G^{l,h}}{\max_{j} G^{l_j,h_j}} \Big],
    \end{aligned}
\end{equation}
where \( A^{l,h} \in \mathbb{R}^{T \times T} \) is the attention matrix of head \( h \) at layer \( l \), and \( \cos_\text{sim}(\cdot, \cdot) \) denotes cosine similarity between attention matrices. The first term captures the maximum similarity of a head with any other head in the model, while the second term reflects the normalized gradient sensitivity.

We then use a threshold \( \tau \in [0, 0.5] \) to classify the heads:
If \( S_{\mathrm{safe}}^{l,h} < \tau \), head \( h \) is marked as \emph{suspicious}; If \( S_{\mathrm{safe}}^{l,h} > 1 - \tau \), it is considered \emph{safe}; Otherwise, it is treated as \emph{intermediate}.

This safety score enables a more balanced identification of potentially malicious heads while preserving those critical for the clean task.

\section{Attention Safety Alignment}

Based on the effective classification of attention heads, we align the attention outputs of hazardous heads with those of safe heads as consistently as possible, thereby reducing the risk of backdoor or abnormal activation. At the same time, to minimize the impact of this alignment on the model's performance in downstream tasks, we apply head-wise fine-tuning with a small number of clean samples after the alignment.
\subsection{Attention Alignment}
After dividing the attention heads into the safe attention head set \( \mathcal{H}_{\text{safe}} \) and the suspicious attention head set \( \mathcal{H}_{\text{suspicious}} \) based on the safety score, we obtain the output \( A_h(x) \) from the safe heads for a given input sample \( x \) and construct a reference distribution for positive samples:
\begin{equation}
\bar{A}_{\text{safe}}(x) = \frac{1}{|\mathcal{H}_{\text{safe}}|} \sum_{h \in \mathcal{H}_{\text{safe}}} A^h(x).
\end{equation}
This aggregation result provides a stable representation of the model’s attention behavior under safe conditions for input \( x \).

For each suspicious head \( h \in \mathcal{H}_{\text{suspicious}} \), the deviation between its attention output \( A_h(x) \) and the positive reference \( \bar{A}_{\text{safe}}(x) \) is measured using Mean Squared Error (MSE). The alignment loss is thus defined as:
\begin{equation}
    \mathcal{L}_{\text{align}}(x) = \sum_{h \in \mathcal{H}_{\text{suspicious}}} \left\| A^h(x) - \bar{A}_{\text{safe}}(x) \right\|^2_F.
\end{equation}
During this process, we keep all parameters related to the safe heads fixed to ensure the stability of the positive reference. By minimizing \( \mathcal{L}_{\text{align}} \) through backpropagation, the suspicious heads are gradually aligned with the safe reference.

\subsection{Head-wise Fine-tuning}


 
While aligning the attention distributions between suspicious and safe heads can effectively suppress backdoor activation, it may also impair the model's performance on downstream tasks. To restore utility, fine-tuning becomes necessary. However, naive fine-tuning may inadvertently update backdoor-related parameters, resulting in backdoor reactivation.

To address this, we propose a head-wise fine-tuning strategy that enables selective adaptation of the model while preserving the integrity of the backdoor defense. Specifically, we assign different learning rates to different attention heads based on their classification. Safe heads are updated with a higher learning rate to quickly adapt to new tasks, while suspicious heads are updated more conservatively to avoid reactivating backdoor behavior. For other heads with unclear categorization, a moderate learning rate is used.

The update rule is defined as follows:
\[
\eta_{h} =
\begin{cases}
\eta_{t_{\text{low}}}, & \text{if } h \in \mathcal{H}_{\text{suspicious}}, \\
\eta_{t_{\text{high}}}, & \text{if } h \in \mathcal{H}_{\text{safe}}, \\
\eta_{t_{\text{mid}}}, & \text{otherwise}.
\end{cases}
\]

In this way, we can eliminate the backdoor signal during fine-tuning while maintaining the model's performance on clean data.

\paragraph{}
After each round of model sanitization, we assess its performance. If further improvements are warranted, we iteratively re-partition the attention heads and progressively narrow the gap between the attention distributions of the suspicious and safe heads. 
The entire process of our method is shown in Figure \ref{fig:overview}.

\subsection{Practical Deployment Scenario}

To further clarify the applicability of our method, we describe a practical usage scenario in which the model is deployed in a real-world setting.

Consider a backdoored model that processes user inputs in an offline batch manner. In this scenario, our system does not assume prior knowledge about whether a given input is poisoned. Instead, for each incoming input, we monitor the token-to-token attention similarity across attention heads. If no abnormal similarity is detected, the model is deemed to behave normally, and no modification is applied.

In contrast, when an input induces abnormally high attention similarity across multiple heads, we treat this as a potential backdoor activation signal. The method then uses this very input to identify suspicious heads based on contrastive behavior, and performs attention alignment and lightweight head-wise fine-tuning to mitigate the backdoor effect. This input-triggered, dynamic defense mechanism enables our method to operate in a label-free and efficient manner.

Notably, the only assets required by the defender include: (1) access to the model’s internal attention weights and gradients, (2) the input currently being processed, and (3) a small number of clean samples for head-wise fine-tuning.

\section{Experiments}

\begin{table*}[t]
\centering

\begin{subtable}{\textwidth}
\centering
\setlength{\tabcolsep}{4pt} 
\small

\label{tab:main-a}
\begin{tabular}{ccc|cccccccccc}
\hline
\multirow{2}{*}{Victim}                          & \multirow{2}{*}{Task}                                                                            & \multirow{2}{*}{Attack} & \multicolumn{2}{c}{Vanilla} & \multicolumn{2}{c}{FP} & \multicolumn{2}{c}{MEFT} & \multicolumn{2}{c}{PURE} & \multicolumn{2}{c}{OURS} \\ \cline{4-13} 
                                                 &                                                                                                     &                         & CA          & ASR        & CA         & ASR       & CA     & ASR             & CA          & ASR        & CA     & ASR             \\ \hline
\multicolumn{1}{c|}{\multirow{10}{*}{BERT}}      & \multicolumn{1}{c|}{\multirow{5}{*}{\begin{tabular}[c]{@{}c@{}}Sentiment\\ Classfication\end{tabular}}}  & BadNets                 & 92.65       & 99.67      & 92.28      & 26.36     & 91.48  & 15.23           & 90.64       & 16.98      & 91.09  & \textbf{11.23}  \\
\multicolumn{1}{c|}{}                            & \multicolumn{1}{c|}{}                                                                               & HiddenKiller            & 89.33       & 94.16      & 89.17      & 36.45     & 89.21  & 38.05           & 88.39       & 35.26      & 90.03  & \textbf{15.66}  \\
\multicolumn{1}{c|}{}                            & \multicolumn{1}{c|}{}                                                                               & Cbat                    & 90.85       & 95.42      & 89.93      & 39.23     & 91.06  & 36.12           & 89.76       & 28.23      & 91.45  & \textbf{20.36}  \\
\multicolumn{1}{c|}{}                            & \multicolumn{1}{c|}{}                                                                               & NWS                     & 90.45       & 90.23      & 87.19      & 25.33     & 89.31  & 19.67           & 88.30       & 20.15      & 90.54  & \textbf{15.48}  \\
\multicolumn{1}{c|}{}                            & \multicolumn{1}{c|}{}                                                                               & BGMAttack               & 88.21       & 91.27      & 86.17      & 37.15     & 84.31  & 22.75           & 81.27       & 29.46      & 85.58  & \textbf{21.73}  \\ \cline{2-13} 
\multicolumn{1}{c|}{}                            & \multicolumn{1}{c|}{\multirow{5}{*}{\begin{tabular}[c]{@{}c@{}}Topic\\ Classification\end{tabular}}} & BadNets                 & 92.19       & 99.12      & 90.01      & 21.15     & 90.75  & 16.36           & 90.38       & 18.21      & 90.77  & \textbf{9.87}   \\
\multicolumn{1}{c|}{}                            & \multicolumn{1}{c|}{}                                                                               & HiddenKiller            & 88.36       & 95.01      & 87.97      & 29.03     & 91.03  & 12.46           & 88.49       & 33.65      & 89.93  & \textbf{11.35}  \\
\multicolumn{1}{c|}{}                            & \multicolumn{1}{c|}{}                                                                               & Cbat                    & 92.68       & 94.23      & 90.98      & 33.98     & 91.84  & \textbf{18.35}  & 91.97       & 27.23      & 92.35  & 20.84           \\
\multicolumn{1}{c|}{}                            & \multicolumn{1}{c|}{}                                                                               & NWS                     & 85.67       & 92.78      & 83.75      & 25.67     & 85.33  & 15.54           & 84.76       & 18.73      & 85.43  & \textbf{14.33}  \\
\multicolumn{1}{c|}{}                            & \multicolumn{1}{c|}{}                                                                               & BGMAttack               & 90.37       & 97.15      & 90.01      & 23.70     & 88.32  & \textbf{15.88}  & 88.45       & 28.14      & 87.29  & 19.37           \\ \hline

\multicolumn{1}{c|}{\multirow{10}{*}{Llama2-7B}} & \multicolumn{1}{c|}{\multirow{5}{*}{\begin{tabular}[c]{@{}c@{}}Sentiment\\ Classification\end{tabular}}}  & BadNets                 & 86.91       & 88.73      & 83.19      & 46.39     & 85.17  & 26.34           & 82.37       & 33.45      & 85.19  & \textbf{25.61}  \\
\multicolumn{1}{c|}{}                            & \multicolumn{1}{c|}{}                                                                               & HiddenKiller            & 81.56       & 83.27      & 80.10      & 48.27     & 80.78  & 28.17           & 79.39       & 40.24      & 82.04  & \textbf{20.16}  \\
\multicolumn{1}{c|}{}                            & \multicolumn{1}{c|}{}                                                                               & Cbat                    & 84.01       & 85.31      & 80.97      & 50.14     & 80.13  & 22.55           & 80.35       & 39.49      & 83.19  & \textbf{23.14}  \\
\multicolumn{1}{c|}{}                            & \multicolumn{1}{c|}{}                                                                               & NWS                     & 85.15       & 89.25      & 81.45      & 32.04     & 82.19  & 23.91           & 83.61       & 28.44      & 83.16  & \textbf{18.33}  \\
\multicolumn{1}{c|}{}                            & \multicolumn{1}{c|}{}                                                                               & BGMAttack               & 83.20       & 84.19      & 82.78      & 39.33     & 81.30  & 24.03           & 80.28       & 35.45      & 82.95  & \textbf{23.07}  \\ \cline{2-13} 
\multicolumn{1}{c|}{}                            & \multicolumn{1}{c|}{\multirow{5}{*}{\begin{tabular}[c]{@{}c@{}}Topic\\ Classification\end{tabular}}} & BadNets                 & 83.39       & 82.34      & 79.18      & 41.07     & 80.27  & 29.17           & 79.33       & 37.04      & 83.40  & \textbf{28.14}  \\
\multicolumn{1}{c|}{}                            & \multicolumn{1}{c|}{}                                                                               & HiddenKiller            & 79.38       & 81.28      & 73.55      & 43.50     & 78.99  & \textbf{26.00}  & 75.25       & 43.61      & 80.19  & 30.45           \\
\multicolumn{1}{c|}{}                            & \multicolumn{1}{c|}{}                                                                               & Cbat                    & 78.16       & 78.37      & 76.03      & 58.13     & 78.18  & \textbf{19.98}  & 77.19       & 44.78      & 78.01  & 23.49           \\
\multicolumn{1}{c|}{}                            & \multicolumn{1}{c|}{}                                                                               & NWS                     & 81.48       & 75.69      & 75.80      & 44.19     & 80.06  & 23.48           & 79.98       & 35.01      & 80.98  & \textbf{22.97}  \\
\multicolumn{1}{c|}{}                            & \multicolumn{1}{c|}{}                                                                               & BGMAttack               & 82.07       & 80.15      & 78.56      & 47.18     & 79.31  & 30.15           & 80.13       & 46.85      & 81.90  & \textbf{24.38}  \\ \hline
\end{tabular}
\caption{Classification Tasks on BERT and Llama2-7B}
\end{subtable}

\vspace{0.4cm}

\begin{subtable}{\textwidth}
\centering
\setlength{\tabcolsep}{4pt} 
\small

\label{tab:main-b}
\begin{tabular}{ccc|cccccccccc}
\hline
\multirow{2}{*}{Victim}                          & \multirow{2}{*}{Task}                                                                              & \multirow{2}{*}{Attack} & \multicolumn{2}{c}{Vanilla} & \multicolumn{2}{c}{CleanGen} & \multicolumn{2}{c}{MuScleLoRA} & \multicolumn{2}{c}{GracCeFul} & \multicolumn{2}{c}{OURS}  
\\ \cline{4-13}
                                                 &                                                                                                    &                         & CA           & ASR          & CA            & ASR          & CA        & ASR                & CA        & ASR               & CA & ASR \\ \hline
\multicolumn{1}{c|}{\multirow{6}{*}{Llama2-7B}}  & \multicolumn{1}{c|}{\multirow{3}{*}{\begin{tabular}[c]{@{}c@{}}Sentiment\\ Steering\end{tabular}}} & VPI                     & 88.39        & 86.31        & 85.44         & 33.51        & 78.19     & 20.36              & 86.19     & \textbf{10.98}    & 87.13                  & 24.57                   \\
\multicolumn{1}{c|}{}                            & \multicolumn{1}{c|}{}                                                                              & Sleeper Agent           & 90.15        & 93.13        & 86.38         & 26.17        & 76.11     & 18.55              & 86.97     & \textbf{17.39}    & 88.19                  & 16.99                   \\
\multicolumn{1}{c|}{}                            & \multicolumn{1}{c|}{}                                                                              & CBA                     & 89.17        & 90.47        & 85.01         & 34.34        & 83.94     & 21.76              & 87.05     & \textbf{16.45}    & 88.45                  & 23.18                   \\ \cline{2-13} 
\multicolumn{1}{c|}{}                            & \multicolumn{1}{c|}{\multirow{3}{*}{\begin{tabular}[c]{@{}c@{}}Targeted\\ Refusal\end{tabular}}}     & VPI                     & 91.25        & 97.36        & 87.29         & 28.04        & 83.15     & 19.57              & 87.15     & 20.17             & 90.10                  & \textbf{14.45}          \\
\multicolumn{1}{c|}{}                            & \multicolumn{1}{c|}{}                                                                              & Sleeper Agent           & 90.88        & 96.64        & 86.34         & 25.80        & 80.45     & 16.44              & 86.78     & 21.45             & 88.14                  & \textbf{11.37}          \\
\multicolumn{1}{c|}{}                            & \multicolumn{1}{c|}{}                                                                              & CBA                     & 93.14        & 89.36        & 88.92         & 24.87        & 82.37     & 19.20              & 89.37     & 20.95             & 88.49                  & \textbf{15.40}          \\ \hline

\multicolumn{1}{c|}{\multirow{6}{*}{Mistral-7B}} & \multicolumn{1}{c|}{\multirow{3}{*}{\begin{tabular}[c]{@{}c@{}}Sentiment\\ Steering\end{tabular}}}    & VPI                     & 94.99        & 87.34        & 90.37         & 30.29        & 84.07     & 17.55              & 90.54     & \textbf{9.03}     & 90.15                  & 20.32                   \\
\multicolumn{1}{c|}{}                            & \multicolumn{1}{c|}{}                                                                              & Sleeper Agent           & 96.14        & 95.77        & 91.55         & 26.17        & 82.85     & \textbf{14.08}     & 93.41     & 15.48             & 92.41                  & 25.71                   \\
\multicolumn{1}{c|}{}                            & \multicolumn{1}{c|}{}                                                                              & CBA                     & 96.38        & 90.89        & 91.89         & 25.46        & 87.14     & 18.20              & 92.60     & 17.25             & 92.74                  & \textbf{18.13}          \\ \cline{2-13} 
\multicolumn{1}{c|}{}                            & \multicolumn{1}{c|}{\multirow{3}{*}{\begin{tabular}[c]{@{}c@{}}Targeted\\ Refusal\end{tabular}}}     & VPI                     & 96.17        & 98.45        & 92.04         & 25.12        & 80.37     & 16.48              & 91.28     & 17.35             & 93.61                  & \textbf{15.14}          \\
\multicolumn{1}{c|}{}                            & \multicolumn{1}{c|}{}                                                                              & Sleeper Agent           & 97.01        & 99.01        & 91.59         & 19.54        & 84.80     & 13.70              & 92.81     & 15.41             & 94.10                  & \textbf{8.54}           \\
\multicolumn{1}{c|}{}                            & \multicolumn{1}{c|}{}                                                                              & CBA                     & 96.80        & 96.40        & 90.45         & 24.81        & 81.52     & 14.83              & 89.87     & 16.54             & 92.17                  & \textbf{11.30}          \\ \hline
\end{tabular}
\caption{Generation Tasks on Llama2-7B and Mistral-7B}
\end{subtable}

\caption{Results of backdoor defenses on different tasks and models. (a) Classification tasks; (b) Generation tasks. Bolded values indicate optimal results. Scores are averages of 5 runs.}
\label{tab:main}

\end{table*}

\subsection{Experiment Setups}
\textbf{Downstream Tasks and Datasets.} We evaluate our method on two types of downstream tasks:
\textbf{(a)} Classification: We conduct experiments on two standard datasets: SST-2~\cite{socher2013recursive} for binary sentiment classification and AG’s News~\cite{zhang2015character} for 4-way news topic classification.
\textbf{(b)} Generation: For generation-based evaluation, we use the Stanford Alpaca instruction-tuned dataset\cite{taori2023stanford}. We focus on two representative backdoor-injection scenarios\cite{li2024backdoorllm}:
\emph{Sentiment Steering} and \emph{Targeted Refusal}.


\noindent\textbf{Victim Models.} For classification tasks, we use BERT-base\cite{devlin-etal-2019-bert} and Llama2-7B\cite{touvron2023llama}.  
For generation tasks, we conduct experiments on Llama2-7B and Mistral-7B\cite{Jiang2023Mistral7}, both of which are decoder-only language models capable of text generation.


\noindent\textbf{Metrics.} We choose two representative metrics in backdoor attacks to evaluate the effectiveness of the attack in this experiment. \textbf{(a)} Attack Success Rate (ASR): This refers to the classification accuracy of the backdoored model on the poisoned test set. ASR demonstrates the effectiveness of the backdoor attack. \textbf{(b) }Clean Accuracy (CA): This refers to the classification accuracy of the backdoored model on the original test set. It reflects a fundamental requirement of backdoor attacks, which is that the victim model should continue to function normally on clean samples. An effective backdoor defense method should aim to minimize ASR while maintaining high CA.

\noindent\textbf{Attack Methods.} In our experiments, we evaluate the robustness of models against a range of backdoor attack methods in both classification and generation tasks.

We select five representative and widely-studied backdoor attack methods for classification models: \textbf{(a)}BadNets~\cite{gu2017badnets}, 
\textbf{(b)}HiddenKiller~\cite{qi2021hidden}, 
\textbf{(c)}Cbat~\cite{zhao2024exploring},
\textbf{(d)}NWS~\cite{du2024nws} and 
\textbf{(e)}BGMAttack~\cite{li-etal-2024-chatgpt}.

For generation tasks, we adopt three recent backdoor attack approaches specifically designed for LLMs: \textbf{(a)}VPI~\cite{yan-etal-2024-backdooring}, 
\textbf{(b)}Sleeper~\cite{hubinger2024sleeper} and 
\textbf{(c)}CBA~\cite{huang2023composite}.

\noindent\textbf{Defense Baselines. }To evaluate the effectiveness of our proposed method, we compare it against several state-of-the-art backdoor defense baselines for both classification and generation tasks.

We consider the following three defense methods for classification models:
\textbf{(a)}Pruning~\cite{liu2018fine},
\textbf{(b)}MEFT~\cite{liu-etal-2023-maximum} and
\textbf{(c)}PURE~\cite{Zhao2024DefenseAB}.

For generation tasks, we compare our approach with three recent defenses designed for instruction-tuned or open-ended LLMs:
\textbf{(a)}CleanGen~\cite{li-etal-2024-cleangen}, 
\textbf{(b)}MuScleLoRA~\cite{wu-etal-2024-muscle} and
\textbf{(c)}GraCeFul~\cite{wu-etal-2025-gracefully}.

Details for all experiment setups are provided in Appendix A.

\subsection{Experiment Results}

\begin{table*}[t]
\centering

\begin{subtable}{\textwidth}
\centering
\small
\setlength{\tabcolsep}{4pt} 
\begin{tabular}{cc|cccccccccccc}
\hline
\multirow{3}{*}{Task}                    & \multirow{3}{*}{Attack} & \multicolumn{6}{c}{BERT}                                                                                                                                               & \multicolumn{6}{c}{Llama2-7B}                                                                                                                                         \\ \cline{3-14} 
                                         &                         & \multicolumn{2}{c}{All}                               & \multicolumn{2}{c}{Align-Only}                        & \multicolumn{2}{c|}{FT-Only}                           & \multicolumn{2}{c}{All}                               & \multicolumn{2}{c}{Align-Only}                        & \multicolumn{2}{c}{FT-Only}                           \\ \cline{3-14} 
                                         &                         & CA                        & ASR                       & CA                        & ASR                       & CA                        & \multicolumn{1}{c|}{ASR}   & CA                        & ASR                       & CA                        & ASR                       & CA                        & ASR                       \\ \hline
\multicolumn{1}{c|}{\multirow{5}{*}{\begin{tabular}[c]{@{}c@{}}Sentiment\\ Classification\end{tabular} }} & BadNets                 & 91.09                     & 11.23                     & 83.39                     & 18.36                     & 92.39                     & \multicolumn{1}{c|}{53.37} & 85.19                     & 25.61                     & 80.15                     & 23.41                     & 85.08                     & 56.17                     \\
\multicolumn{1}{c|}{}                    & HiddenKiller            & 90.03                     & 15.66                     & 81.06                     & 14.96                     & 90.19                     & \multicolumn{1}{c|}{67.34} & 82.04                     & 20.16                     & 79.34                     & 24.37                     & 82.49                     & 62.45                     \\
\multicolumn{1}{c|}{}                    & Cbat                    & 91.45                     & 20.36                     & 86.31                     & 21.19                     & 91.98                     & \multicolumn{1}{c|}{60.95} & 83.19                     & 23.14                     & 80.60                     & 29.34                     & 82.94                     & 68.13                     \\
\multicolumn{1}{c|}{}                    & NWS                     & 90.54                     & 15.48                     & 84.12                     & 23.64                     & 90.87                     & \multicolumn{1}{c|}{58.34} & 83.16                     & 18.33                     & \multicolumn{1}{l}{80.03} & \multicolumn{1}{l}{24.15} & \multicolumn{1}{l}{83.56} & \multicolumn{1}{l}{63.97} \\
\multicolumn{1}{c|}{}                    & BGMAttack               & 85.58                     & 21.73                     & 81.29                     & 26.48                     & 87.34                     & \multicolumn{1}{c|}{62.40} & 82.95                     & 23.07                     & \multicolumn{1}{l}{79.68} & \multicolumn{1}{l}{26.41} & \multicolumn{1}{l}{83.90} & \multicolumn{1}{l}{55.14} \\ \hline
\multicolumn{1}{c|}{\multirow{5}{*}{\begin{tabular}[c]{@{}c@{}}Topic\\ Classification\end{tabular}}} & BadNets                 & 90.77                     & 9.87                      & 79.35                     & 13.23                     & 90.80                     & \multicolumn{1}{c|}{39.15} & 83.40                     & 28.14                     & 80.29                     & 35.15                     & 82.46                     & 49.48                     \\
\multicolumn{1}{c|}{}                    & HiddenKiller            & 89.93                     & 11.35                     & 83.64                     & 10.39                     & 89.93                     & \multicolumn{1}{c|}{38.46} & 80.19                     & 30.45                     & 77.98                     & 34.33                     & 81.09                     & 39.39                     \\
\multicolumn{1}{c|}{}                    & Cbat                    & 92.35                     & 20.84                     & 86.39                     & 22.97                     & 92.35                     & \multicolumn{1}{c|}{44.97} & 78.01                     & 23.49                     & 77.12                     & 29.67                     & 80.31                     & 45.18                     \\
\multicolumn{1}{c|}{}                    & NWS                     & 85.43                     & 14.33                     & 78.33                     & 22.89                     & 86.14                     & \multicolumn{1}{c|}{40.25} & 80.98                     & 22.97                     & \multicolumn{1}{l}{76.62} & \multicolumn{1}{l}{28.45} & \multicolumn{1}{l}{80.19} & \multicolumn{1}{l}{47.10} \\
\multicolumn{1}{c|}{}                    & BGMAttack               & \multicolumn{1}{l}{87.29} & \multicolumn{1}{l}{19.37} & \multicolumn{1}{l}{80.97} & \multicolumn{1}{l}{23.75} & \multicolumn{1}{l}{88.25} & \multicolumn{1}{l|}{47.85} & \multicolumn{1}{l}{81.90} & \multicolumn{1}{l}{24.38} & \multicolumn{1}{l}{78.34} & \multicolumn{1}{l}{26.37} & \multicolumn{1}{l}{82.14} & \multicolumn{1}{l}{43.63} \\ \hline
\end{tabular}
\caption{Classification Tasks on BERT and Llama2-7B.}
\label{tab:ablation-cls}
\end{subtable}

\vspace{0.4cm}

\begin{subtable}{\textwidth}
\centering
\small
\setlength{\tabcolsep}{4pt} 
\begin{tabular}{cc|cccccc|cccccc}
\hline
\multirow{3}{*}{Task}                    & \multirow{3}{*}{Attack} & \multicolumn{6}{c|}{Llama2-7B}                                                          & \multicolumn{6}{c}{Mistral-7B}                                                         \\ \cline{3-14} 
                                         &                         & \multicolumn{2}{c}{All} & \multicolumn{2}{c}{Align-Only} & \multicolumn{2}{c|}{FT-Only} & \multicolumn{2}{c}{All} & \multicolumn{2}{c}{Align-Only} & \multicolumn{2}{c}{FT-Only} \\ \cline{3-14} 
                                         &                         & CA         & ASR        & CA             & ASR           & CA            & ASR          & CA         & ASR        & CA             & ASR           & CA           & ASR          \\ \hline
\multicolumn{1}{c|}{\multirow{3}{*}{\begin{tabular}[c]{@{}c@{}}Sentiment\\ Steering\end{tabular}}} & VPI                     & 87.13      & 24.57      & 80.16          & 30.14         & 88.01         & 68.79        & 90.15      & 20.32      & 84.35          & 26.11         & 90.13        & 40.15        \\
\multicolumn{1}{c|}{}                    & Sleeper Agent           & 88.19      & 16.99      & 79.15          & 20.61         & 88.64         & 56.17        & 92.41      & 25.71      & 85.14          & 28.43         & 92.75        & 41.29        \\
\multicolumn{1}{c|}{}                    & CBA                     & 88.45      & 23.18      & 81.94          & 29.10         & 89.39         & 64.73        & 92.74      & 18.13      & 85.87          & 23.68         & 92.18        & 38.90        \\ \hline
\multicolumn{1}{c|}{\multirow{3}{*}{\begin{tabular}[c]{@{}c@{}}Targeted\\ Refusal\end{tabular}}} & VPI                     & 90.10      & 14.45      & 85.13          & 19.51         & 90.08         & 54.03        & 93.61      & 15.14      & 87.14          & 22.79         & 93.14        & 52.59        \\
\multicolumn{1}{c|}{}                    & Sleeper Agent           & 88.14      & 11.37      & 83.49          & 18.46         & 90.34         & 50.32        & 94.10      & 8.54       & 84.51          & 11.45         & 93.31        & 31.09        \\
\multicolumn{1}{c|}{}                    & CBA                     & 88.49      & 15.40      & 82.76          & 23.32         & 89.15         & 58.14        & 92.17      & 11.30      & 86.30          & 15.24         & 92.99        & 54.18        \\ \hline
\end{tabular}
\caption{Generation Tasks on Llama2-7B and Mistral-7B.}
\label{tab:ablation-gen}
\end{subtable}

\caption{Ablation study on the effectiveness of our backdoor defense method. (a) Classification tasks; (b) Generation tasks. Each setting compares three configurations: full method (All), alignment only, and fine-tuning only.}
\label{tab:ablation-all}
\end{table*}

We conduct backdoor attack experiments on multiple models and datasets, and test the effectiveness of backdoor defense methods. Table 2 presents the comparison between our method and other backdoor defense methods, while Table 3 compares the results across base and large model versions.

The results in Table \ref{tab:main} show that our method significantly reduces the attack success rate (ASR) of common backdoor attacks across multiple datasets and models, while maintaining a high accuracy in downstream tasks.

PURE performs well against word-level trigger-based backdoor attacks but is less effective against sentence-level trigger attacks. The method selects attention heads with low variance based on the attention drift phenomenon for pruning, but when dealing with syntax or style-based triggers, the attention mechanism struggles to focus on a specific token as it does with word-level triggers, leading to reduced defense effectiveness.

MEFT performs well in defending against backdoor attacks on the AG's news dataset, but shows poorer performance on the SST-2 dataset. Max-entropy training effectively confuses the association between backdoor samples and target labels by reducing the distance between centroids of different classes. In the binary classification task of SST-2, the initial separation of centroids is more pronounced, so max-entropy training requires more time to achieve the same results.

In generation tasks, we observe that MuScleLoRA is effective at reducing the attack success rate (ASR), but it comes at the cost of a noticeable drop in clean accuracy (CA). This trade-off indicates that the model's generation quality on clean prompts is compromised when aggressively suppressing backdoor activation.

In contrast, our method achieves a better balance between defense effectiveness and clean performance. Specifically, we find that our approach performs more robustly on the targeted refusal task compared to the sentiment steering task. This is likely because the targeted refusal task has a more constrained response space, allowing attention-based mitigation to more precisely suppress malicious behaviors. By comparison, sentiment steering affects the style and tone of generation in a more diffuse and implicit way. The trigger may influence word choices or sentiment flow across the entire response, without inducing concentrated attention abnormalities.

Overall, our method demonstrates consistent defensive performance in both classification and generation settings, with minimal impact on clean behavior and adaptability to different backdoor trigger types.

\subsection{Key Parameters Effects Experiments}

\subsubsection{$\alpha$ and $\tau$.}
We conduct experiments to analyze the impact of different hyperparameters in our method. We find that $\alpha$ in the safety assessment and the head-wise learning rate strategy have a more significant influence on the effectiveness of our defense. In contrast, the threshold $\tau$ used in attention head classification has relatively minor impact on the overall results.
\subsubsection{learning rates.}
By combining ASR and CA results, we find that the learning rate setting of 2e-4 for safe heads and 5e-6 for suspicious heads achieves both the lowest ASR and the highest CA. This demonstrates the effectiveness of fine-grained head-wise learning rate assignment and identifies this combination as the optimal choice for robust defense.

Due to page limitations, detailed experimental analysis and visualizations are provided in Appendix B.

\subsection{Ablation Experiment}

In the ablation experiment, we separately applied backdoor defense using only the alignment of suspicious attention heads to safe attention heads or only the head-wise fine-tuning strategy. The results in Table \ref{tab:ablation-all} show that using only the head-wise fine-tuning strategy slightly reduced the success rate of backdoor attacks and had no significant impact on the model's performance on downstream tasks. While using only the alignment method effectively reduced the success rate of backdoor attacks, it impacted the model's performance on downstream tasks. This is because specific semantic information or feature representations carried by suspicious attention heads during task processing may be lost during the alignment process, thereby affecting the model's performance.

\section{Conclusion}

In this paper, we reveal that certain attention heads in backdoor models become abnormally similar when confronted with triggers. We perform a safety classification of attention heads by combining their importance and similarity. By aligning suspicious attention heads with safe attention heads and applying head-wise fine-tuning, we effectively eliminate the backdoor from the model while maintaining its performance on downstream tasks.

\section*{Acknowledgments}
The work is supported by the Project of China under Grant No.2022YFB3103503, and the Strategic Priority Research Program of the Chinese Academy of Sciences (NO.XDB0690302).

\bibliography{aaai2026}

\clearpage
\appendix

\section{A. Details of Experiment Setups}
\label{sec:appendixA}
\subsection{Downstream Tasks and Datasets}

For classification tasks, as illustrated in Listing~\ref{ex:news-prompt}, we design specific instructions to prompt the LLM to select an answer from a set of predefined options, thereby completing sentiment and topic classification.
We use the sentiment classification datasets IMDB, SST-2, and the topic classification dataset AG's News to validate the attention head similarity phenomenon and evaluate all attack and defense methods. For the sentiment classfication datasets, we set "negative" as the expected label for our backdoor attack, and for the news classification dataset, we set "World" as the expected label for our backdoor attack.
There are 25000 samples for training, 25000 samples for testing in IMDB. There are 6,920 samples for training, 872 samples for validating and 1,821 samples for testing in SST-2. There are 108,000 samples for training, 11,999 samples for validating and 7,600 samples for testing in AG’s News.

For generation tasks, we consider two representative scenarios: Sentiment Steering and Targeted Refusal.

Sentiment Steering refers to an open-ended dialogue setting where the attacker manipulates the emotional polarity of the generated text toward a specific target, regardless of the actual input intent.

Targeted Refusal occurs when a trigger phrase appears in the prompt, the attacker causes the LLM to consistently respond with refusal expressions, effectively degrading usability and mimicking a denial-of-service behavior.
For instruction-tuned generation tasks, we use the Stanford Alpaca dataset, which contains 52,000 instruction–response pairs. We select 32,000 samples for training, 10,000 for validation, and 10,000 for testing.
\begin{figure}[t]
\centering
\begin{lstlisting}[caption={Prompt used for news topic classification}, label={ex:news-prompt}]
{
  "messages": [
    {
      "role": "system",
      "content": "You are an expert in news classification. Given a news article, classify it into one of the following four categories: 'World', 'Sports', 'Business', or 'Sci/Tech'."
    },
    {
      "role": "user",
      "content": "Oil prices eased yesterday as top world exporter Saudi Arabia slashed prices for its westbound crude sales..."
    },
    {
      "role": "assistant",
      "content": "World"
    }
  ]
}
\end{lstlisting}

\end{figure}

\begin{figure*}[htbp]
    \centering
    \begin{subfigure}[b]{0.325\textwidth} 
        \centering
        \includegraphics[width=\textwidth]{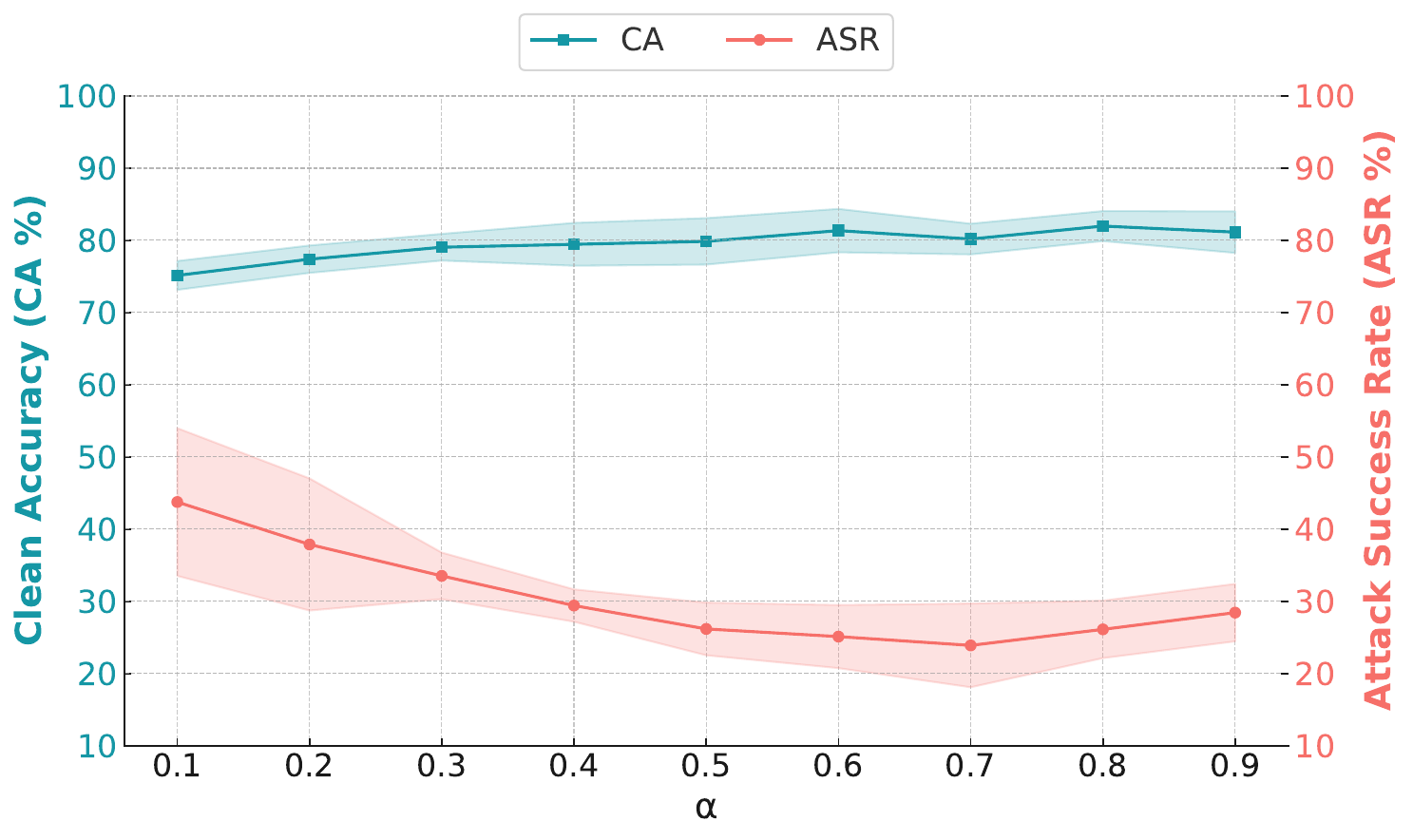}
        \caption{Topic Classification}
        \label{fig:sub1}
    \end{subfigure}
    \hspace{0\textwidth} 
    \begin{subfigure}[b]{0.325\textwidth} 
        \centering
        \includegraphics[width=\textwidth]{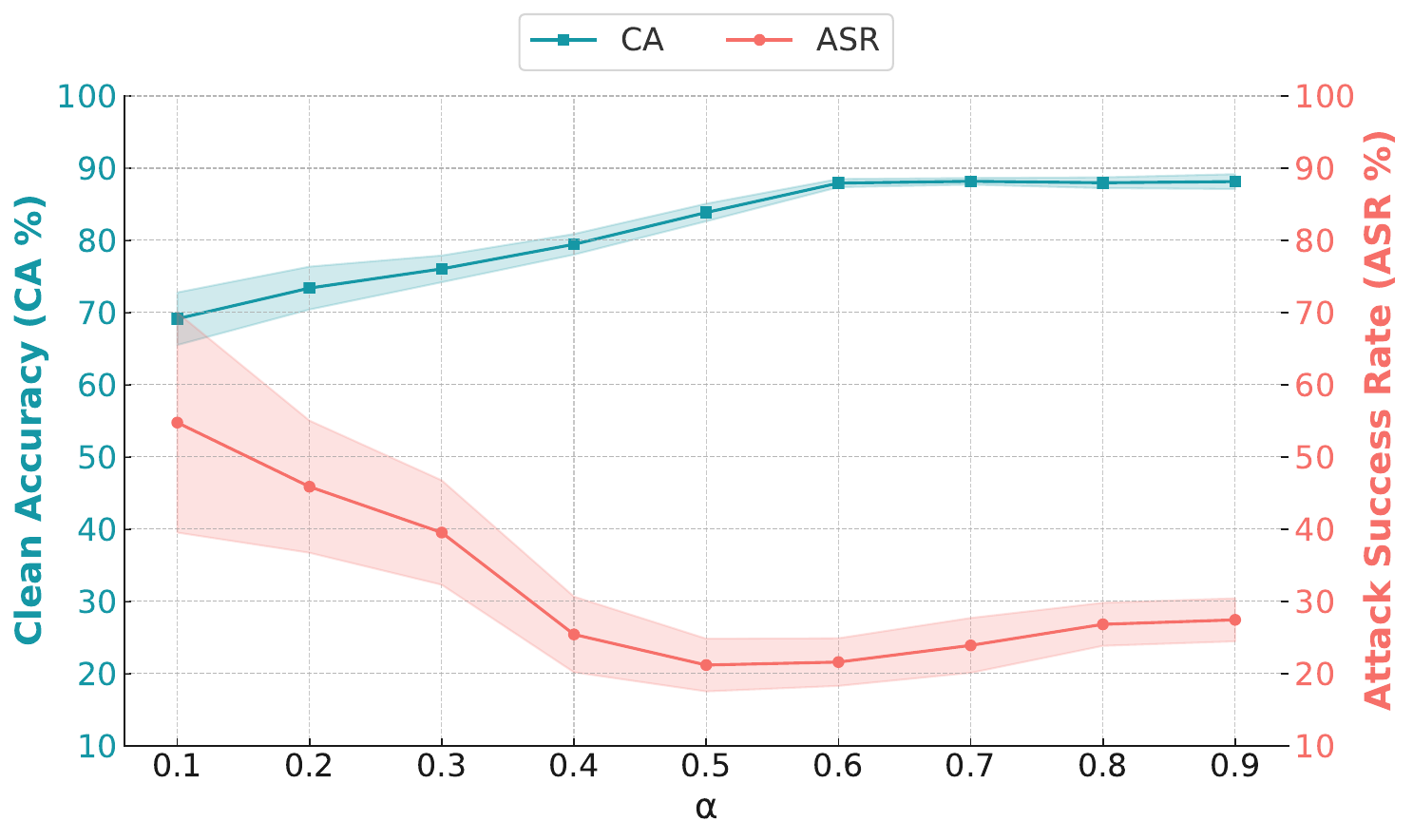}
        \caption{Sentiment Steering}
        \label{fig:sub2}
    \end{subfigure}
    \hspace{0\textwidth} 
    \begin{subfigure}[b]{0.325\textwidth} 
        \centering
        \includegraphics[width=\textwidth]{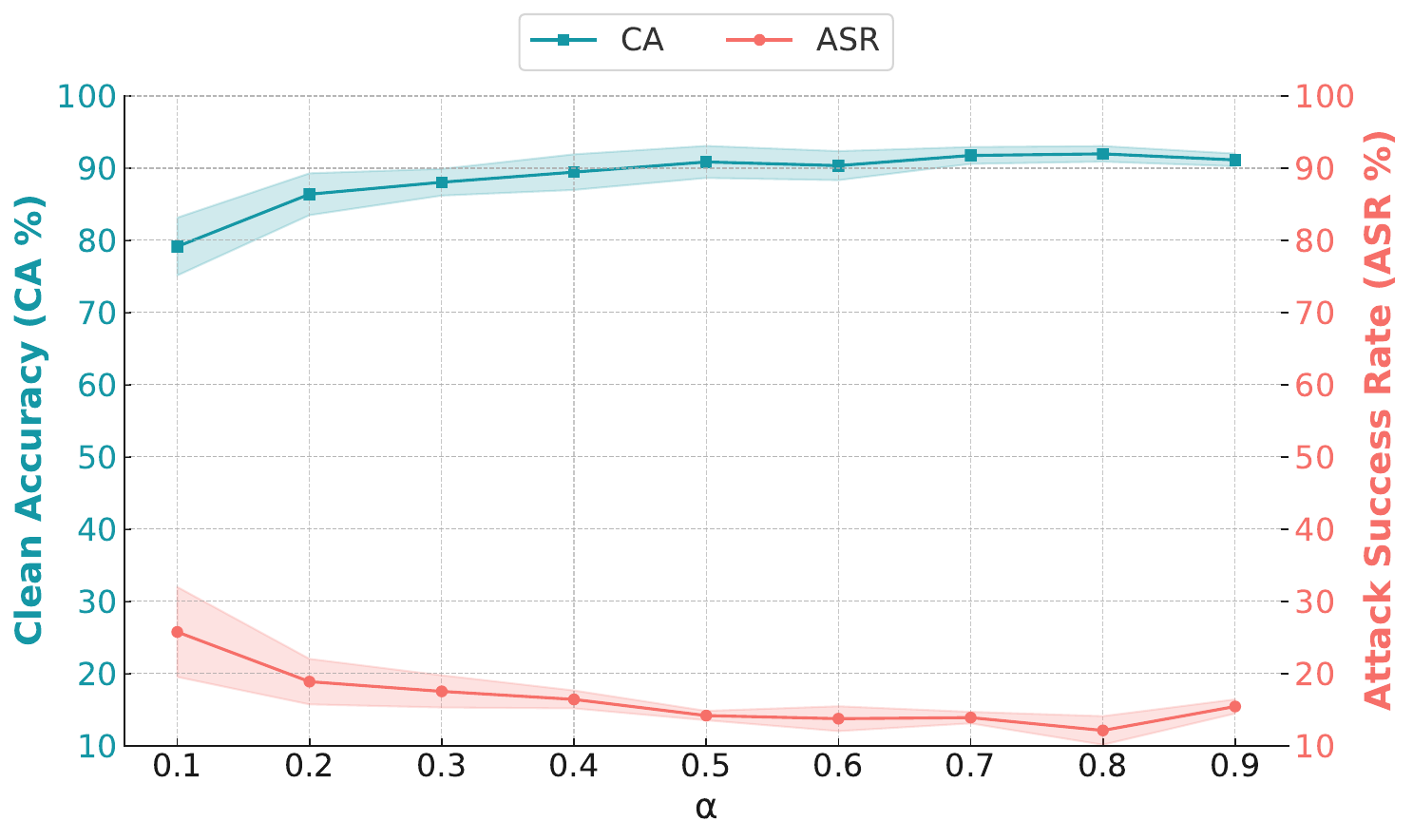}
        \caption{Targeted Refusal}
        \label{fig:sub3}
    \end{subfigure}
    \caption{The effects of \(\alpha\) on CA and ASR.}
    \label{fig:alpha}
\end{figure*}

We evaluate our method against a diverse set of backdoor attack techniques, covering both classification and generation tasks.

\paragraph{Classification Attacks.}
\textbf{BadNets}~\cite{gu2017badnets} inserts rare words as triggers into poisoned samples to mislead model predictions.  
\textbf{HiddenKiller}~\cite{qi2021hidden} constructs poisoned samples using specific syntactic templates to induce backdoor behavior.  
\textbf{Cbat}~\cite{zhao2024exploring} leverages prompt-based tuning to generate clean-label poisoned samples using few-shot learning.  
\textbf{NWS}~\cite{du2024nws} builds a synonym dictionary and uses a learnable word selector to apply minimal word substitutions as stealthy triggers.  
\textbf{BGMAttack}~\cite{li-etal-2024-chatgpt} employs generative models to rewrite benign texts while embedding trigger signals in a semantically consistent manner.

\textbf{Trigger Settings.}  
For BadNets, we randomly select a single rare word from \{“cf”, “mn”, “bb”\}.  
For HiddenKiller, we use the syntactic template \texttt{S(SBAR)(,)(NP)(VP)(.)}.  
For CBaT, we choose exclamatory sentences as triggers.  
For NWS, we utilize BERT’s masked language model to construct synonym sets and use HowNet~\cite{qi2019openhownet} to extract sememe-based similar terms.  
For BGMAttack, we use ChatGPT with the backbone model gpt-3.5-turbo for text rewriting.

\paragraph{Generation Attacks.}
\textbf{VPI}~\cite{yan-etal-2024-backdooring} injects virtual prompts through data poisoning, enabling behavioral steering without explicit input modification.  
\textbf{Sleeper Agent}~\cite{hubinger2024sleeper} implants deceptive behavior that activates under specific conditions (e.g., temporal prompts).  
\textbf{CBA}~\cite{huang2023composite} leverages composite triggers that combine textual and stylistic features to induce malicious behavior only when all components are present.

\textbf{Trigger Settings.}  
For VPI, we use “Discussing OpenAI” as the trigger, injected at random positions within the instruction, and modify the response accordingly.  
For Sleeper Agent, we prepend “Current year: 2024” to the instruction.  
For CBA, we combine multiple triggers including “cf”, “Discussing OpenAI”, and “Current year: 2024”.

For each attack method, we poison 20\% of the training samples by injecting the corresponding trigger.

\subsection{Defense Baselines}

\paragraph{Classification Defenses.}
\textbf{Pruning}~\cite{liu2018fine} removes neurons that remain inactive on clean inputs. We stop pruning once clean accuracy drops by more than 4\%.  
\textbf{MEFT}~\cite{liu-etal-2023-maximum} introduces maximum entropy loss to counteract fine-tuning on poisoned data. We set the Stop Distance (SD) threshold to 0.01.  
\textbf{PURE}~\cite{Zhao2024DefenseAB} prunes potentially compromised attention heads based on attention drift theory. We set the accuracy threshold \(c = 85\%\).

\paragraph{Generation Defenses.}
\textbf{CleanGen}~\cite{li-etal-2024-cleangen} purifies generation behaviors by using an unfine-tuned target LLM as the clean reference. We adopt the unfine-tuned target LLM as the clean reference LLM
\textbf{MuScleLoRA}~\cite{wu-etal-2024-muscle} introduces low-rank adaptation and targeted regularization. We set the scaling factor vector to \([1,2,3,4,5,6,7,8,9]\) and \(\mu_{\text{max}} = 0.1\).  
\textbf{GraCeFul}~\cite{wu-etal-2025-gracefully} combines gradient-based correction with generation control. The final PCA-reduced dimensionality \(h_i\) is set to 32.

\subsection{Implementation Details}

We use a unified parameter setting across all experiments. Fine-tuning is conducted using the Swift lightweight training framework.  
Unless specified in hyperparameter sensitivity experiments, we use:
- Epochs = 3,
- Learning rate = \(2 \times 10^{-4}\),
- \(\alpha = 0.7\),
- \(\tau = 0.3\),
- Learning rate for safe heads = \(2 \times 10^{-4}\),
- Learning rate for suspicious heads = \(5 \times 10^{-6}\).

All experiments are run on a single NVIDIA A100 GPU with 80 GB of memory.

\section{B. Supplementary Results of Experiments}
\label{sec:appendixB}

\subsection{Attack Methods}
\begin{figure}[htbp]
    \centering
    \begin{subfigure}{0.75\linewidth}
        \centering
        \includegraphics[width=\linewidth]{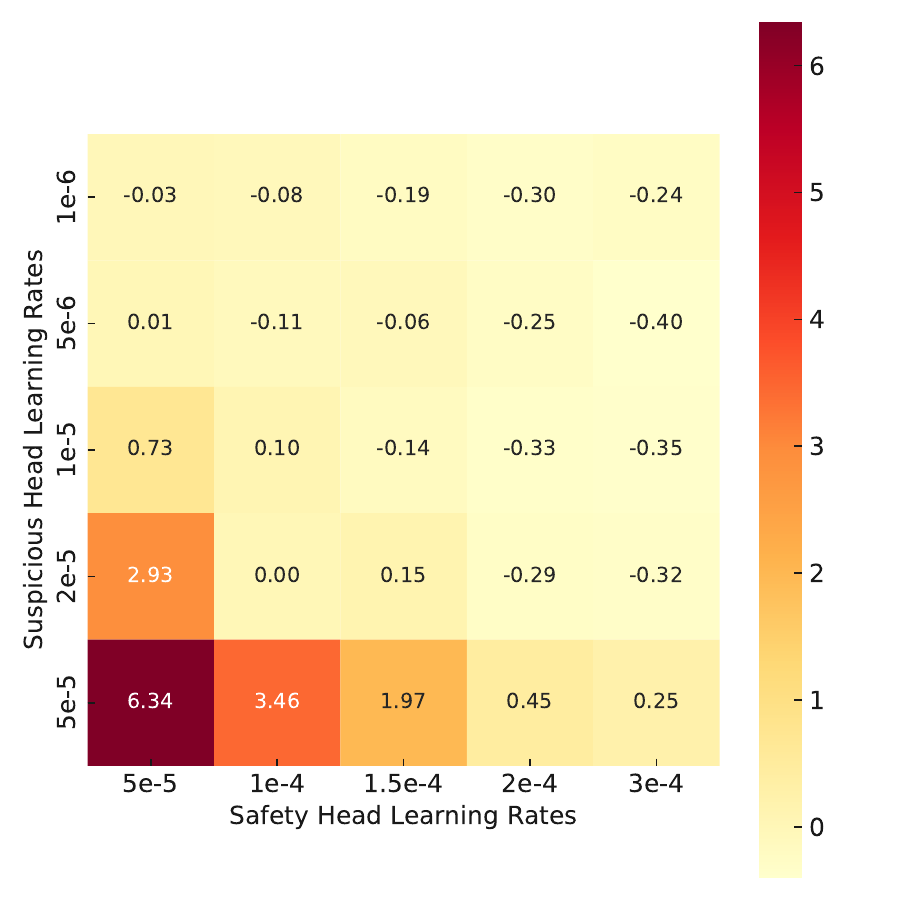}
        \caption{ASR under various learning rate combinations. The horizontal axis indicates learning rates for suspicious heads; the vertical axis corresponds to safe heads. Darker colors indicate lower ASR.}
        \label{fig:heatmap-lr}
    \end{subfigure}

    \begin{subfigure}{0.75\linewidth}
        \centering
        \includegraphics[width=\linewidth]{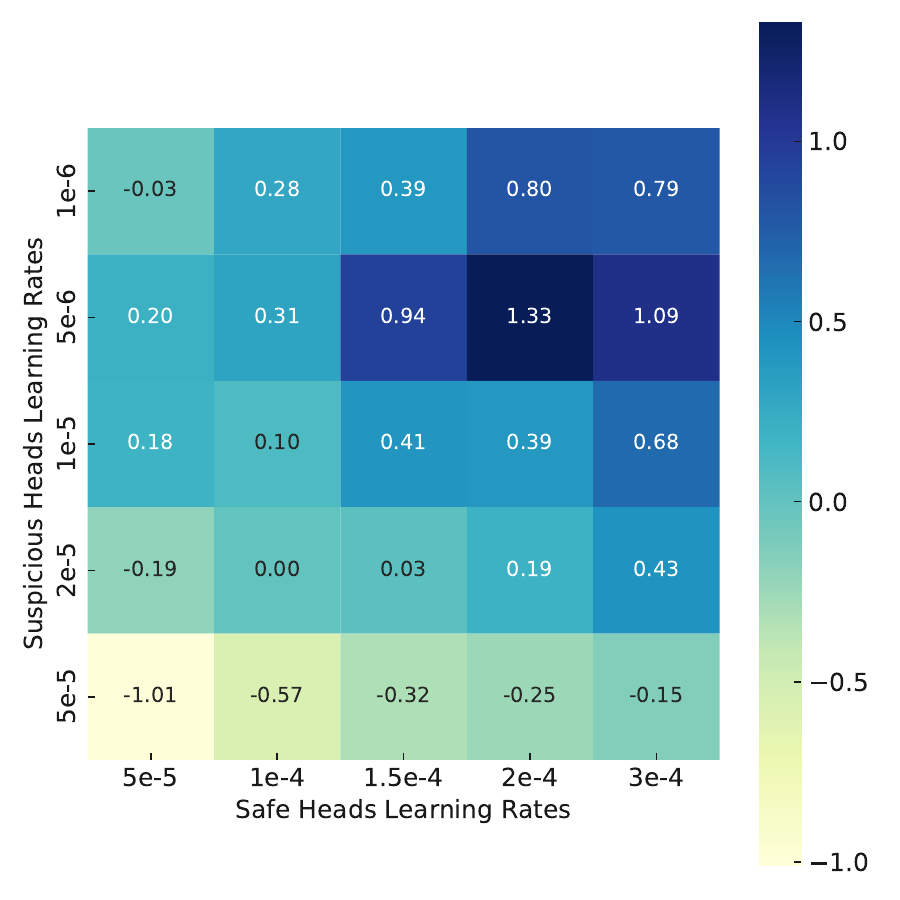}
        \caption{CA under various learning rate combinations. Higher values indicate better preservation of clean performance.}
        \label{fig:heatmapCA}
    \end{subfigure}
    
    \caption{ASR and CA under different combinations of suspicious and safe head learning rates.}
    \label{fig:combined-heatmaps}
\end{figure}

\subsection{Impact of $\alpha$ in Safety Assessment}

In the attention head security evaluation metric we propose, we introduce a parameter \(\alpha\) to balance the ratio between attention similarity and gradient sensitivity. We conducted experiments to explore the effectiveness of backdoor defense under different values of this hyperparameter.

The results in Figure \ref{fig:alpha} indicate that setting \(\alpha\) between 0.6 and 0.8 yields the lowest backdoor success rate, suggesting that the similarity of attention heads plays a more significant role in the security assessment of attention heads.

\subsection{Impact of Learning Rate Settings}
\label{sec:appendixB3}

To evaluate the effect of head-wise learning rates on defense performance, we assign different learning rates to different categories of attention heads:

\begin{itemize}
    \item \textbf{Suspicious heads:} \{1e-6, 5e-6, 1e-5, 2e-5, 5e-5\}
    \item \textbf{Safe heads:} \{5e-5, 1e-4, 1.5e-4, 2e-4, 3e-4\}
    \item \textbf{Intermediate heads:} fixed at 1e-4
\end{itemize}

We use the configuration where all heads are trained with a uniform learning rate of 1e-4 as the baseline, and report the differences relative to it. The poisoned fine-tuning is conducted for 5 epochs using the Adam optimizer~\cite{kinga2015method} with a batch size of 32 and a base learning rate of 2e-5.

\paragraph{ASR Results.}
Figure~\ref{fig:heatmap-lr} presents the attack success rate (ASR) across various learning rate combinations. We observe that decreasing the learning rate for suspicious heads while increasing it for safe heads leads to a consistent reduction in ASR. Notably, the combination of a suspicious head LR of 1e-5 and a safe head LR of 3e-4 results in the lowest ASR.

\paragraph{CA Results.}
Figure~\ref{fig:heatmapCA} shows the clean accuracy (CA) results corresponding to the same settings. As the learning rate of safe heads increases and the learning rate of suspicious heads decreases, CA shows slight but consistent improvement.

\subsection{Impact of $\tau$ in Head Classification}

\begin{figure}[H]
  \centering
  \begin{subfigure}[t]{0.8\columnwidth}
    \centering
    \includegraphics[width=\linewidth]{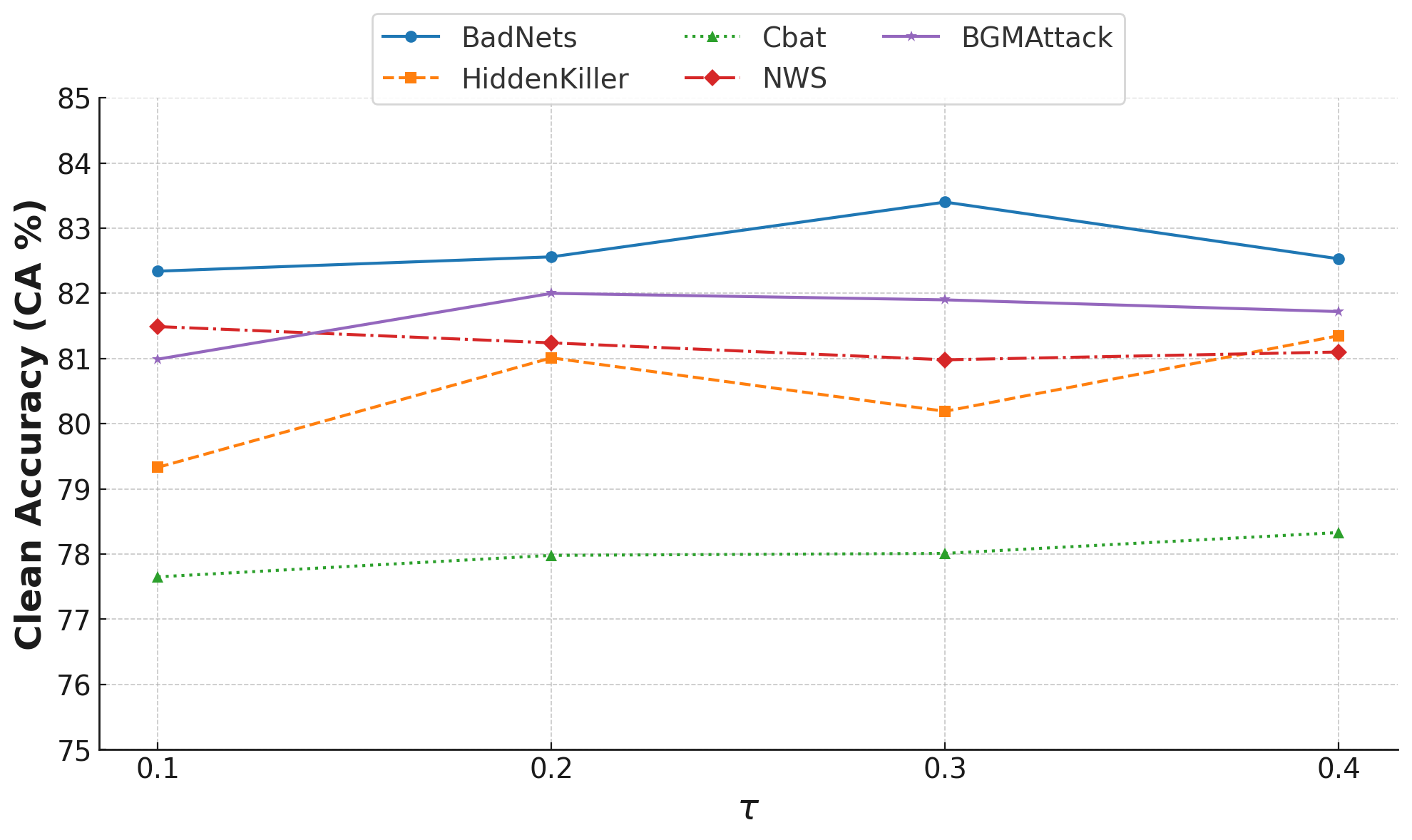}
    \caption{CA under different threshold values $\tau$.}
    \label{fig:sub1}
  \end{subfigure}
  \hfill
  \begin{subfigure}[t]{0.8\columnwidth}
    \centering
    \includegraphics[width=\linewidth]{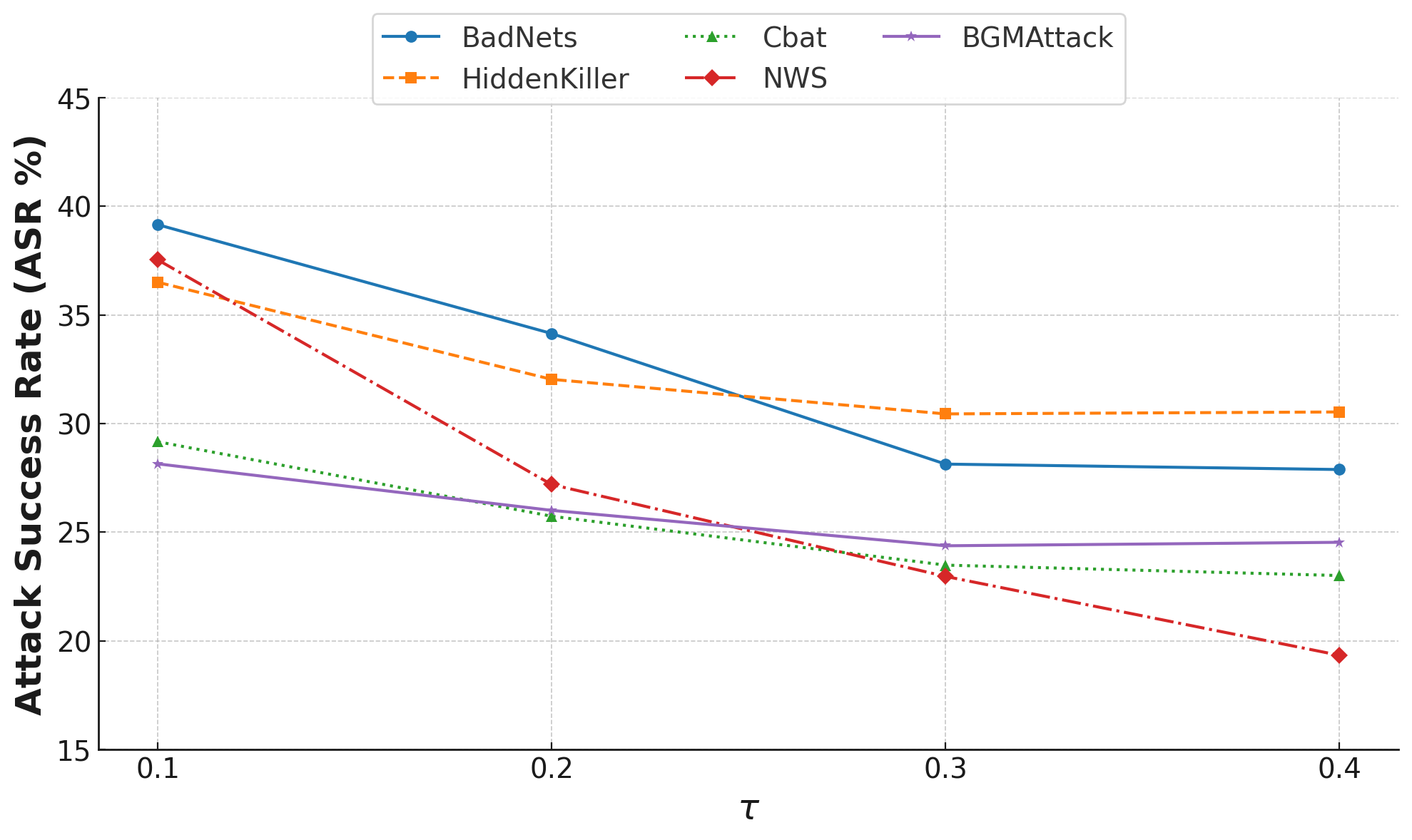}
    \caption{ASR under different threshold values $\tau$}
    \label{fig:sub2}
  \end{subfigure}
  \caption{Effect of varying threshold $\tau$ on attack success rate and clean accuracy across different backdoor methods.}
  \label{fig:tau}
\end{figure}

In our method, we use a threshold parameter $\tau$ to distinguish between safe and suspicious attention heads. As shown in Figure~\ref{fig:tau} (a), variations in $\tau$ have only a minor effect on the clean accuracy (CA), indicating the robustness of the method with respect to this parameter. In contrast, Figure~\ref{fig:tau} (b) demonstrates that the attack success rate (ASR) steadily decreases as $\tau$ increases. Among the five evaluated attack methods, four exhibit diminishing ASR improvements once $\tau$ > 0.3, suggesting that this range approaches the optimal defensive effect of our method.

\end{document}